\documentclass[11pt,aps,preprint,tightenlines,groupadress]{revtex4}
\usepackage{graphicx}

\def\etal{\mbox{et al.}}

\begin{document}




\title{Microscopic approach to the bcc phase of solid $^{\bf 4}$He
}

\author{R. Rota and J. Boronat$^{\ast}$\thanks{$^\ast$Corresponding author. 
Email: jordi.boronat@upc.edu
\vspace{6pt}}
\\ 
\vspace{6pt}  
{\em{Departament de F\'\i sica i Enginyeria Nuclear, Campus Nord B4-B5,
\\
Universitat Polit\`ecnica de Catalunya, E-08034 Barcelona, Spain}}
\\
 }

\begin{abstract}
The coexistence between the hcp and bcc phases of solid $^4$He at fixed
pressure, and at three different temperatures, is studied by means of the
path integral Monte Carlo method. Microscopic results for the main
energetic and structure properties of the two phases are reported. The
differences between hcp and bcc $^4$He are shown to be small with the
obvious exception of the static structure factor.  When crossing the phase
transition line, the most significant changes are observed in the kinetic
energy per  particle and in the Lindemann ratio, both pointing to a less
correlated quantum solid in the bcc side. No off-diagonal long-range order
is observed in the hcp and bcc perfect lattices.

\end{abstract}

\maketitle

\section{Introduction}

Quantum solids have been studied for many years in the context of low-temperature 
condensed-matter physics. 
Contrarily to classical solids, the high zero-point motion of the atoms
makes their oscillation around the lattice sites significant even at very low
temperature. This makes quantum crystals
have large Lindemann ratios and unambiguous anharmonic effects. 
Solid helium in its two stable isotopes $^4$He and $^3$He, and their
mixtures, represent the best example of quantum solid~\cite{glydebook}.
Helium has a mass so small and its atomic interaction is so
shallow that remains liquid even in the limit of zero temperature. It is
necessary to pressurize $^4$He and $^3$He up to $\sim$ 25 and 30 bar,
respectively, to freeze them  at $T=0$. Because of the different mass 
and their different quantum statistics ($^4$He is a boson and $^3$He is a
fermion), the solidification is in the two cases different: $^4$He
crystallizes into an hcp solid whereas $^3$He does it into a bcc one. The
phase diagram of both systems offers also the counterpart: $^3$He evolves
to an hcp phase at higher pressures and $^4$He presents a very narrow bcc
region at temperatures $T \sim 1.5-1.7$ K. 

The theoretical and experimental interest in solid $^4$He has been
reinforced since the experimental observation in 2004 of non classical rotation
of inertia (NCRI) in torsional oscillator experiments by Kim and
Chan~\cite{kim}. This
experiment, subsequent ones by the same team, and independent measures in a
number of different laboratories have confirmed these findings but the
fraction of mass decoupled (superfluid fraction) depends strongly on the
quality of the crystal~\cite{balibar}. The experimental data and their interpretation as
unambiguous signatures of the existence of the supersolid state 
are still matter of debate. The observation by Beamish and
collaborators~\cite{beamish} of
an increase of the shear modulus of hcp $^4$He at nearly the same onset
temperature  for NCRI ($T_0 \simeq 100$ mK) has lead to think that, at least in a partial way,
 the
NCRI signal can be attributed to the change in the elastic properties.
However, there is a general consensus that not all the torsional oscillator
results can be interpreted solely on plasticity terms~\cite{maris,kim2}.

\begin{figure}[t]
\begin{center}
\includegraphics[width=0.7\linewidth,angle=-90]{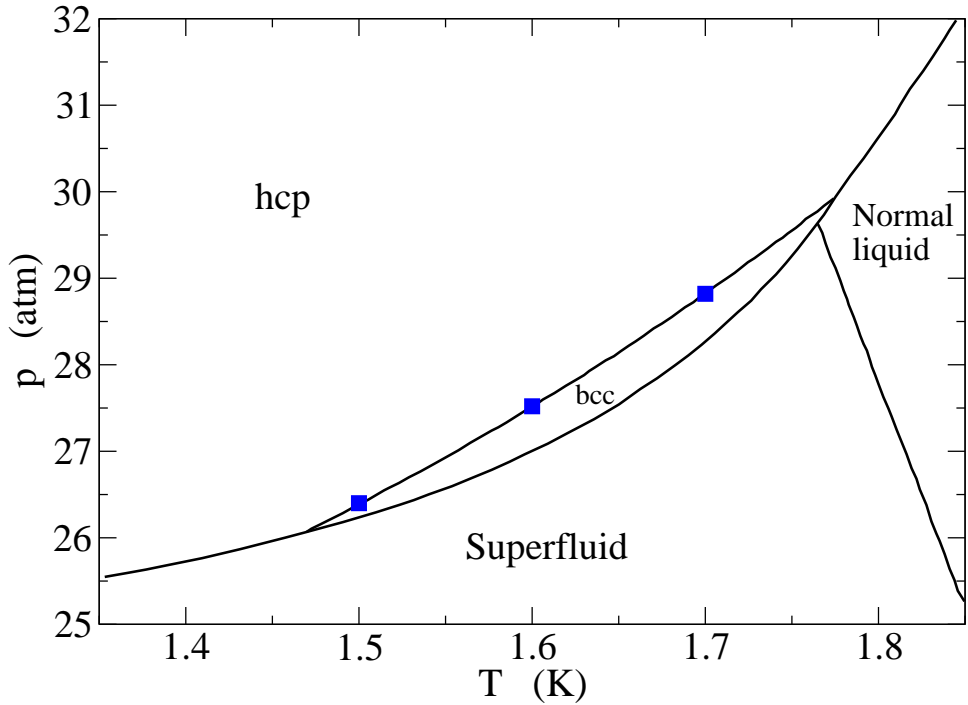}
\caption{(Color online) Pressure-temperature phase diagram in the stable region of the bcc
phase of solid $^4$He. The squares correspond to the thermodynamic points
studied in the present work. 
}
\label{figurepd}
\end{center}
\end{figure}

The main focus of this recent research activity has been hcp $^4$He since
this is the stable phase at very low temperatures. However, Eyal
\etal~\cite{eyal} \ have
reported recently torsional oscillator experiments on bcc $^4$He at
temperatures between 1.3 and 1.9 K that show similar phenomena to those
measured in the hcp phase below 100 mK. This is a surprising result since,
even if the disorder of the sample is eventually large,
the temperatures of this experiment are one order of magnitude larger than
in hcp. This is not the only relevant result related to the bcc phase that
emerged in the last years. Some time ago, Markovich \etal~\cite{markovich} \ reported
neutron-scattering measurements of phonons in bcc $^4$He and found an
unexpected ``opticlike" mode along the $[110]$ direction. This new mode has
been theoretically interpreted by Gov~\cite{gov} in terms of correlated dipolar
interactions. 

The boundaries of the bcc phase of $^4$He were first determined by Vignos
and Fairbank~\cite{vignos} in 1961 after analyzing their measurements of the velocity of
longitudinal sound. The existence of this phase was confirmed after x-ray
scattering by Schuch and Mills~\cite{schuch}. Later on, Grilly and
Mills~\cite{grilly} measured the
volume change along the hcp-bcc coexistence line and Edwards and
Pandorf~\cite{edwards}
performed accurate measurements of the heat capacity. Updated accounts of
the thermodynamic properties of bcc $^4$He can be found in Refs.
\cite{hoffer,grilly2}. In Fig. \ref{figurepd}, the $P$-$T$ phase diagram in the region of
interest is plotted. As one can see, the stability of the bcc solid is only
$\sim 0.04$ K wide at fixed pressure and it represents an intermediate
phase between the hcp solid and the superfluid in the range of
temperatures  $T\simeq 1.5$-$1.75$ K.

From the theoretical point of view, the high anharmonicity of solid $^4$He
makes its analysis much more demanding since the usual harmonic theory used in
classical solids is not applicable. The self-consistent phonon (SCP) theory incorporates
approximately the short-range correlations induced by the interatomic
potential, leading to reasonable descriptions of many of its
properties~\cite{glydebook}.
However, a more accurate study of quantum solids requires of a
microscopic approach. To this end, quantum Monte Carlo (QMC) methods have been widely applied
in the past to the study of solid $^4$He. However, the main part of the QMC
work is devoted to the hcp phase since it is the stable crystal in the limit of
very low temperature. Results on the bcc phase are scarce and concentrated
to the limit of zero temperature~\cite{pederiva,pederiva2, galli}. To our knowledge, 
the only existing work
in the bcc stable $P$-$T$ region shown in Fig. \ref{figurepd} was done by
Ceperley~\cite{ceperley} with the main goal of estimating the momentum distribution and
kinetic energies after some neutron scattering experiments~\cite{blasdell}. Our aim in the
present work is to carry out a QMC study of the bcc phase at three
temperatures and densities corresponding to the experimental points shown
in Fig. \ref{figurepd} and compare them with the corresponding hcp ones
along the coexistence line. As we will show, the differences between both
phases are found small except for the kinetic energies and Lindemann ratios
where the less correlated character of the bcc lattice is revealed.         

The rest of the paper is organized as follows. In Sec. II, we introduce the
path integral Monte Carlo (PIMC) method used in our simulations. Sec. III
comprises the main results obtained, and finally a brief account of the main
conclusions is contained in Sec. IV.

\section{Theoretical approach}

The structure and energetic properties of the hcp-bcc coexistence line have
been studied using the path integral Monte Carlo (PIMC) method.
PIMC provides a fundamental approach
in the study of strongly interacting quantum systems at finite
temperature by means of a stochastic estimation of the
thermal density matrix~\cite{CeperleyRev}. As it is well known, the partition function
\begin{equation}
\label{PartitionFunction} 
Z = \rm{Tr}(e^{-\beta\hat{H}}) =
\int dR \langle R \vert e^{-\beta\hat{H}} \vert R \rangle 
\end{equation}
allows for a full microscopic description of the properties of a given
system with Hamiltonian $\hat{H} = \hat{K} + \hat{V}$ at a temperature $T =
(k_B\beta)^{-1}$ (we use the position basis $\vert R
\rangle = \vert {\bf r}_1, \ldots {\bf r}_N \rangle$, with $N$
the number of particles). However, in quantum systems the noncommutativity of the 
kinetic  and 
potential energy operators (respectively, $\hat{K}$ and $\hat{V}$) makes
impractical a direct calculation of $Z$ using its definition
(\ref{PartitionFunction}).

The basic idea of PIMC is to use the convolution property of the thermal
density matrix $\rho(R,R';\beta)=\langle R \vert e^{-\beta\hat{H}} \vert R'
\rangle$, in order to rewrite the partition function as
\begin{equation}
\label{Convolution} 
Z = \int \prod_{i=0}^{M-1}dR_i \,
\rho(R_i,R_{i+1};\varepsilon) \ , 
\end{equation} 
with $\varepsilon=\beta/M$
and the boundary condition $R_M = R_0$. For sufficiently large $M$, we
recover the high-temperature limit of the thermal density matrix, where it
is legitimate to separate the kinetic contribution from the potential one
(Primitive Approximation). If one ignores the quantum statistics of 
particles, the distribution law appearing in Eq. \ref{Convolution} is
positive definite and can be interpreted as a probability distribution
function which can be sampled by standard metropolis Monte Carlo methods.

In practice, the PIMC method consists in mapping the finite-temperature
quantum system onto a classical system made up of closed ring polymers. This
technique may be referred as an ``exact" method, in the sense that using an
accurate approximation for the high-temperature density matrices, the
results are not affected by this approximation within the statistical
error. However, the number $M$ of convolution
terms (beads) necessary to reach the convergence of Eq. \ref{Convolution}
to the exact value of $Z$ is inversely proportional to the temperature of
the system. This means that, when approaching the interesting quantum
regime at very low temperature, $M$ increases fast making simulations hard,
if not impossible, due to the very low efficiency in the sampling of the
long chains involved.
To overcome this problem, it is important to work out high-order
approximation schemes for the density matrix, able to work with larger
values of $\varepsilon$. The approximation we use in this work is called
Chin Approximation (CA)~\cite{Sakkos09}. CA is based on a fourth-order
expansion of the operator $e^{-\beta\hat{H}}$ which makes use of the double
commutator $[[\hat{V},\hat{K}],\hat{V}]$, this term being related to the
gradient of the interatomic potential. With respect to the Takahashi-Imada
approximation~\cite{TakahashiImada}, which is accurate to fourth order only
for the trace, the new feature appearing in the CA is the presence of
coefficients weighting the different terms in the expansion of the action.
These coefficients are continuously tunable, making possible to force the
error terms of fourth order to roughly cancel each other and get an
effective sixth-order approximation.

An additional problem one has to deal with when simulating quantum
many-body systems with PIMC arises from the indistinguishable nature of 
particles. In the case of bosons like $^4$He, the indistinguishability of
particles does not change the positivity of the probability distribution in
Eq. \ref{Convolution} and the symmetry of $\rho(R,R';\beta)$ can be
recovered via the direct sampling of permutations between the ring polymers
representing the quantum particles. To this purpose, we use the Worm Algorithm 
(WA)~\cite{BoninsegniWorm} which samples very efficiently the permutation
space. The
basic idea of this technique is to work in an extended configuration space,
given by the union of the ensemble $Z$, formed by the usual closed-ring
configurations, and the ensemble $G$, which is made up of configurations
where all the polymers but one are closed. Thanks to the presence of an
open polymer, one can search the atoms involved in a permutation
cycle by means of single particle updates, which do not suffer of a low
acceptance rate and guarantee an efficient and ergodic sampling of the
bosonic permutations. We have to notice that the probability distribution
used to sample the configurations in $G$ is not equal to the one appearing
in Eq. \ref{Convolution} and, therefore, these configurations cannot be
used to calculate diagonal properties, such as the energy or the superfluid
density. However, the $G$-configurations can be used to compute
off-diagonal observables such as the one-body density matrix $\rho_1({\bf
r}_1,{\bf r}_1')$. Furthermore, the WA samples both diagonal
and off-diagonal configurations and then it is able to give an estimation of the
normalization factor of $\rho_1({\bf r}_1,{\bf r}_1')$. In this way, one
can compute the properly normalized one-body density matrix and so 
avoid the systematic uncertainties introduced by an a posteriori
normalization factor.

\section{Results}

We have carried out our PIMC study of the hcp-bcc coexistence line
particularizing the simulations in the three experimental thermodynamic
points shown in Fig. \ref{figurepd}. Explicitly, in the $P$-$T$ phase diagram
the coordinates of 
these points are A$=(1.5,26.419)$, B$=(1.6,27.572)$, and C$=(1.7,29.029)$
with units of bar and kelvin for the pressure and temperature,
respectively. The corresponding densities are taken from the experimental
data contained in Refs.~\cite{hoffer,grilly2}: ($\rho_A^{\text hcp} = 0.028834$, $\rho_A^{\text
bcc} = 0.028571$), ($\rho_B^{\text hcp} = 0.028954$, $\rho_B^{\text
bcc} = 0.028679$), and ($\rho_C^{\text hcp} = 0.029080$, $\rho_C^{\text
bcc} = 0.028805$), all in units of $\AA^{-3}$. 

In the Hamiltonian of the system the potential part is built as a sum of
pair-wise interatomic potentials, $\hat{V}=\sum_{i<j}^N V(r_{ij})$, with
$V(r)$ of Aziz type~\cite{aziz}. Simulations are performed using periodic boundary
conditions with a number of atoms per simulation cell of $N=180$ (hcp) and
$N=128$ (bcc); tail corrections to the energy due to the use of a finite number of
particles are estimated by running simulations with different $N$ values and
extrapolating its linear behavior in $1/N$ to the thermodynamic limit $1/N
\rightarrow 0$~\cite{claudi}. The convergence to the limit $\varepsilon \rightarrow 0$ is
achieved with relatively large values of $\varepsilon$ ($\varepsilon = 0.010$ K$^{-1}$)
due to the high accuracy of the CA~\cite{Sakkos09}.

\begin{table}
\caption{PIMC results of the total ($E/N$) and partial energies ($V/N$, $K/N$) for the three
thermodynamic points in the hcp-bcc coexistence line here analyzed.
Experimental values of the bcc energies, reported in the first row, are taken
from Ref.~\cite{hoffer}. Figures in parenthesis stand for the statistical errors.}
 {\begin{tabular}{lcccccc} \toprule
  & \multicolumn{2}{c}{A ($T = 1.5$ K)} & \multicolumn{2}{c}{B ($T = 1.6$
  K)} & \multicolumn{2}{c}{C ($T = 1.7$ K)} \\
  & bcc & hcp & bcc & hcp & bcc & hcp \\
  \colrule
  & & & & & & \\
  $E/N$ expt. (K) & -5.95 & & -5.93 & & -5.91 & \\  
  $E/N$ PIMC (K) & -6.166(7) & -6.058(7) & -6.137(8) & -6.025(7) & -6.091(9) & -5.984(6) \\
  & & & & & & \\ 
  $V/N$ (K) & -30.103(3) & -30.448(3) & -30.230(3) & -30.605(3) & -30.372(3) & -30.752(2) \\
  $K/N$ (K) & 23.936(5) & 24.391(7) & 24.093(5) & 24.580(5) & 24.281(7) & 24.768(5) \\
\botrule
\end{tabular}}
\end{table}

PIMC results for the total energy per particle in the three mentioned
thermodynamic points are reported in Table 1. The difference between the
energies of the two lattices in the coexistence line is small but, in the
three cases, the bcc energies are larger than the hcp ones (in absolute
values). This difference is mainly due to the slightly smaller density in the
bcc side with respect to the hcp one, with a smaller effect of the lattice
type (at the same density). On the other hand, PIMC energies lie below the
available experimental energies of the bcc solid in an amount $\sim 0.20$
K, a feature already observed in liquid $^4$He and attributable to the
particular Aziz potential used in the present simulations~\cite{boro}.

In Table 1, we also have included the partial energies per particle. The
results for the kinetic energy per particle $K/N$ are the most relevant
because it is possible to measure them using deep inelastic neutron
scattering. As it is well known, if the momentum of the incoming neutron is
high enough the impulse approximation holds and the atomic momentum
distribution $n(k)$ is attainable, and from it the kinetic energy. Blasdell
\etal~\cite{blasdell} reported experimental results of $K/N$ for both solid lattices: at
$T=1.07$ K and hcp $K/N=23.6$ K, and at $T=1.72$ K and bcc $K/N=23.7$ K (the
densities of both phases were the same). If just the classical estimation
$3/2 kT$ is considered for an isothermical comparison, one can see that the
kinetic energy of the bcc crystal is significantly smaller. PIMC results
by Ceperley~\cite{ceperley} also agree with this trend (in this case the comparison is
made between fcc and bcc lattices). Our present results allow for a best
comparison since they are obtained following the coexistence line. As one
can see in Table 1, there is a nearly constant decrease of $\sim 0.5$ K
when going from the hcp point to the bcc one and our results are in overall
agreement with the experimental determinations of Blasdell
\etal~\cite{blasdell}. The
thermal effects in the kinetic energy for any of the two lattices are
dominated by the classical term  $3/2 kT$ that gives a difference of 0.15 K
from one point to the next one.

\begin{figure}[t]
\begin{center}
\includegraphics[width=0.7\linewidth,angle=-90]{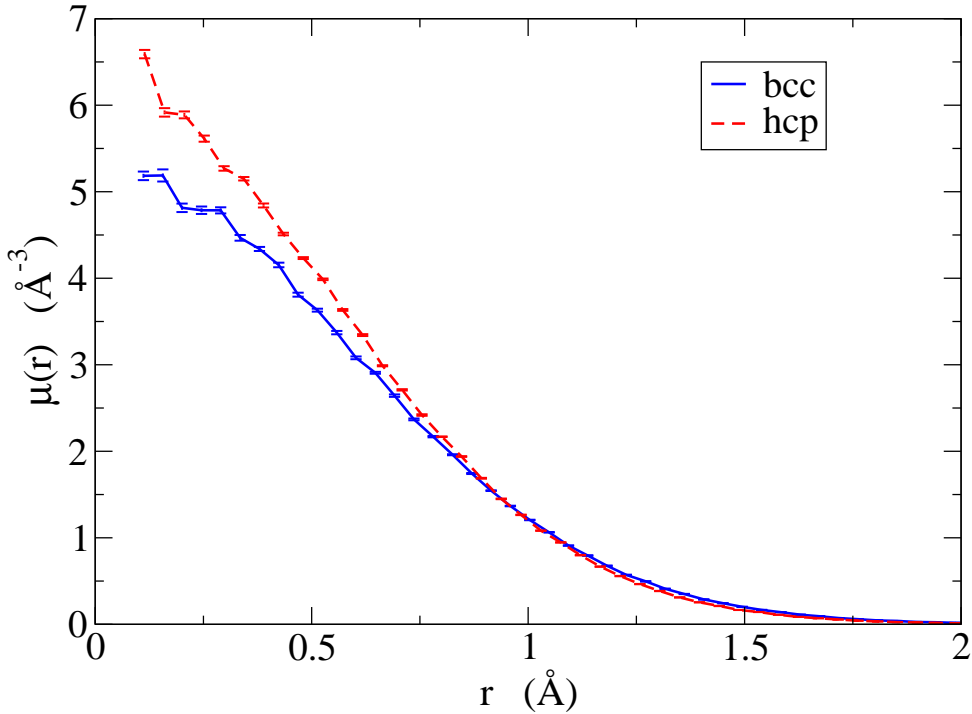}
\caption{(Color online) Density profile $\mu(r)$ of $^4$He atoms around the lattice sites for
the two lattices at $T=1.5$ K. 
}
\label{figuremu}
\end{center}
\end{figure}

We have also calculated some structural properties of the bcc phase to
better characterize it and to identify specific signatures of this
lattice symmetry. One of the most relevant in the microscopic study of any
solid phase is the density profile $\mu(r)$, defined as the probability of
finding a particle in the interval $(r,r+dr)$ around any of the lattice
points of the crystal. Results of $\mu(r)$ at the thermodynamic point A and
for the two coexistence lattices are shown in Fig. \ref{figuremu}. As
expected, the density profile of the bcc crystal is a bit wider than the
hcp one, with a decrease of localization as it corresponds to its slightly
more open structure. From the density profile it is possible to calculate
the mean squared displacement around a site,
\begin{equation}
\langle {\bf u}^2 \rangle= 4 \pi \int_0^{\infty} dr \, r^4 \mu(r) \  ,
\label{u2}
\end{equation} 
and from it to estimate the Lindemann ratio,
\begin{equation}
\gamma = \frac{\sqrt{\langle {\bf u}^2 \rangle}}{a} \ ,
\label{lindemann}
\end{equation}
$a$ being the nearest-neighbor distance in the perfect crystalline lattice.
We have obtained $\gamma_{\text{hcp}}=0.26$ and $\gamma_{\text{bcc}}=0.28$,
in agreement with experimental data~\cite{glydebook}. The Lindemann ratio is a good measure
of the zero-point motion and thus the larger value of $\gamma$ 
for the bcc lattice points to an enhancement of its quantum nature.

\begin{figure}[t]
\begin{center}
\includegraphics[width=0.7\linewidth,angle=-90]{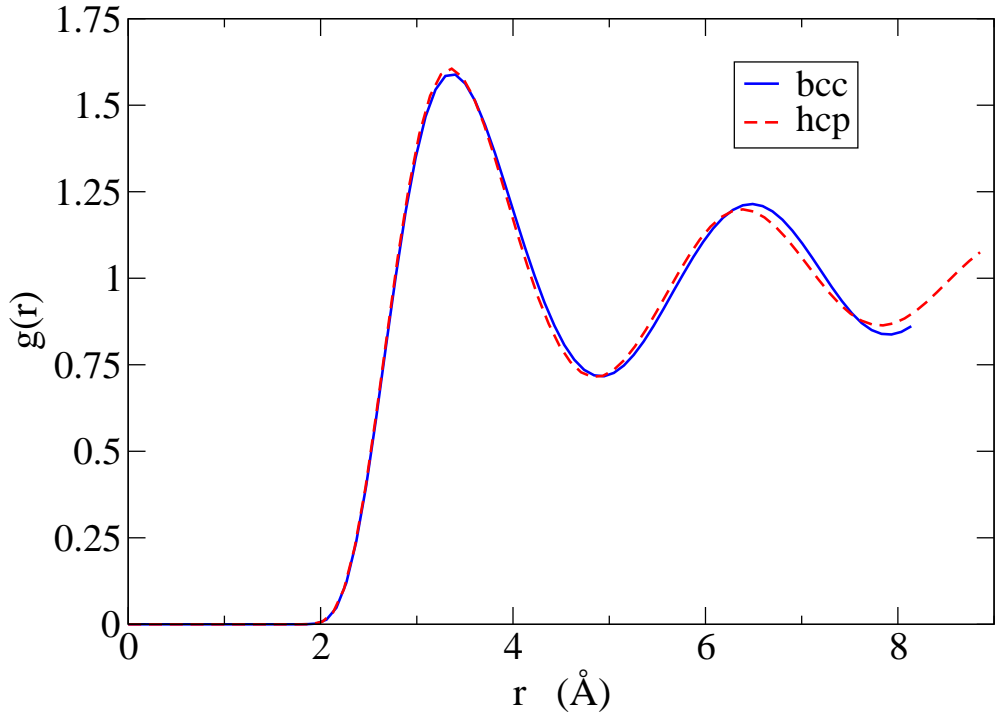}
\caption{(Color online) Two-body radial distribution function for the hcp
and bcc phases of solid $^4$He at $T=1.5$ K. 
}
\label{figuregr}
\end{center}
\end{figure}

\begin{figure}[]
\begin{center}
\includegraphics[width=0.7\linewidth,angle=-90]{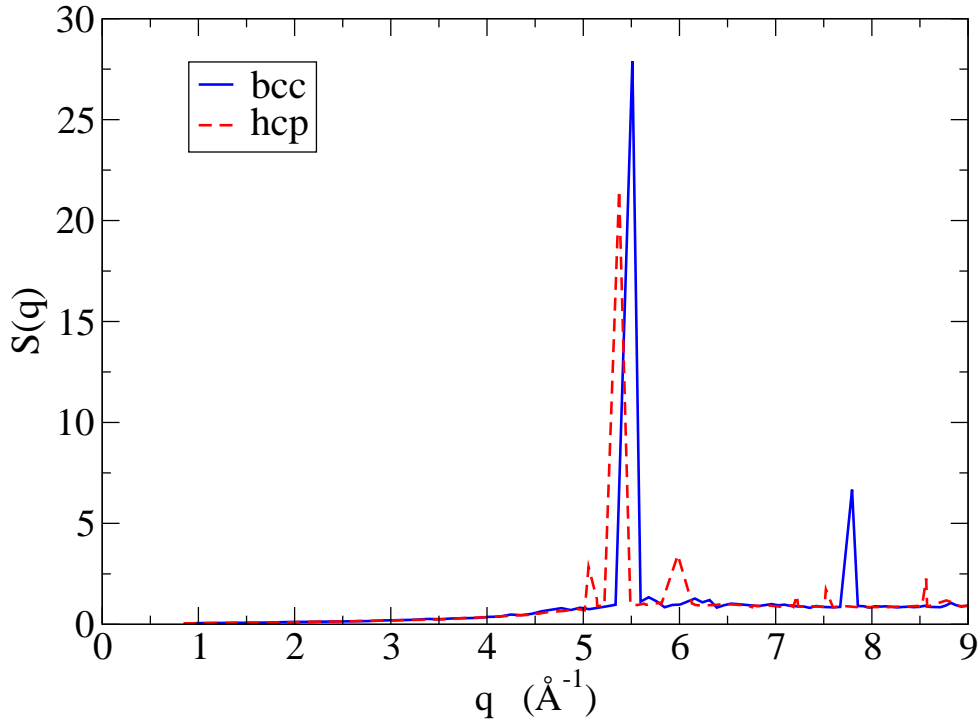}
\caption{(Color online) Static structure factor for the hcp
and bcc phases of solid $^4$He at $T=1.5$ K. 
}
\label{figuresk}
\end{center}
\end{figure}

In Fig. \ref{figuregr}, we show results of the two-body radial distribution
function for both lattices in the coexistence at $T=1.5$ K. One can see the
usual oscillations due to the periodic spatial order and only a tiny
difference between both lattices. The height of the main peak of the bcc
$g(r)$ is slightly smaller than the hcp one but this is more an effect of its
smaller density than a consequence of the lattice symmetry. The difference
in the spatial order induced by the lattice is more clear in reciprocal
space by defining the static structure factor
\begin{equation}
S(q) = \frac{1}{N} \left\langle \sum_{i=1}^N e^{-i {\bf q} \cdot {\bf r}_i}
\, \sum_{j=1}^N e^{i {\bf q} \cdot {\bf r}_j} \right\rangle  \ .
\label{eseq}
\end{equation}

Results for $S(q)$ at both sides of the coexistence line at $T=1.5$ K are
shown in Fig. \ref{figuresk}. Several Bragg peaks are clearly identified;
they are located at the expected points for each type of lattice. As it is
well known from solid state physics, the determination of the static
structure factor is the best way for identifying the particular lattice of
the crystal and, in fact, the x-ray measures by Schuch and
Mills~\cite{schuch} were the
definite prove of the existence of a stability region for bcc $^4$He.

\begin{figure}[]
\begin{center}
\includegraphics[width=0.7\linewidth,angle=-90]{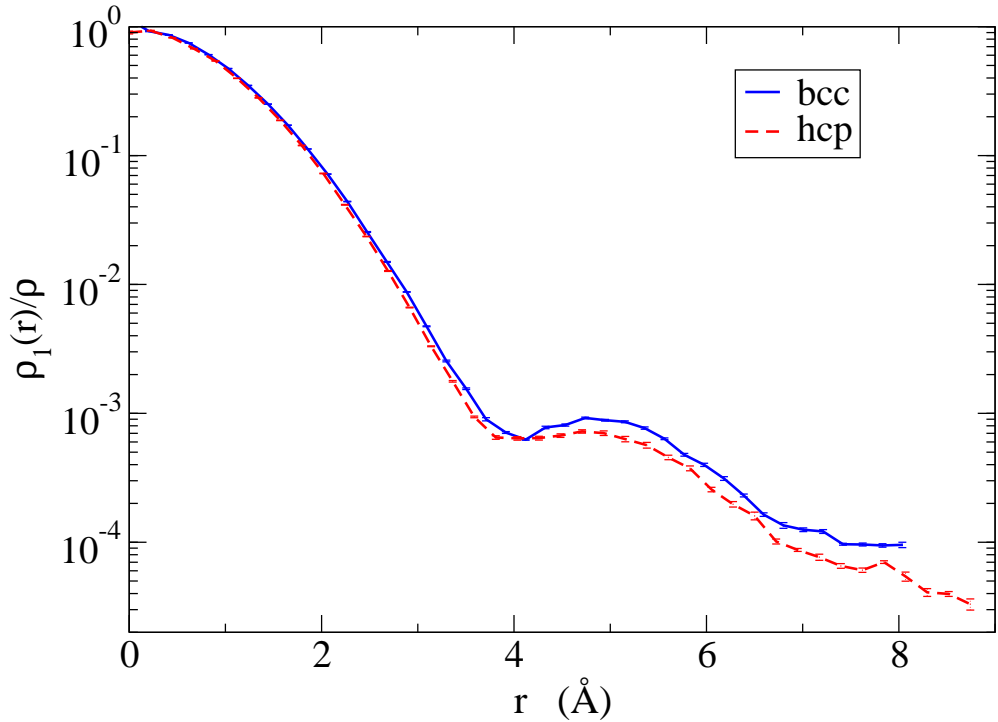}
\caption{(Color online) One-body density matrix for the hcp
and bcc phases of solid $^4$He at $T=1.5$ K. 
} 
\label{figurerho}
\end{center}
\end{figure}

Much of our present interest in the bcc phase of solid $^4$He lies on the
unexpected results found by Eyal \etal~\cite{eyal} which point to disorder-induced
mobility, that manifests in the TO experiments as the typical mass
decoupling observed in hcp at very low temperatures ($T < 200$ mK). This
could open the possibility of a supersolid scenario at higher temperatures.
Any signature of supersolidity would appear in off-diagonal properties of
the system, like the one-body density matrix $\rho_1(r)$ defined as  
\begin{equation}
\frac{\rho_1({\bf r}_1,{\bf r}_1^\prime)}{\rho}  = \frac{1}{Z} \, \int d{\bf
r}_2,\ldots,{\bf r}_N \ \rho({\bf r}_1,\ldots,{\bf r}_N;{\bf
r}_1^\prime,\ldots,{\bf r}_N;\beta) \ .
\label{obdm}
\end{equation}
With the help of the WA one can calculate this function with high
efficiency and avoid normalization problems that were present in other
permutation sampling methods. In Fig. \ref{figurerho}, we show results for  $\rho_1(r)$ 
at $T=1.5$ K calculated for the two lattices. The shape of the two results
is very similar and coincides with previous estimations: a fast decay up to
$\sim 4\ \AA$, a small increase near the position of the first neighbor, and
finally a kind of exponential decay to zero.  The most
relevant feature of these results is the absence of off-diagonal long-range
order, which would manifest as a nonzero asymptotic constant (condensate fraction)  when 
$r \rightarrow \infty$. Therefore, neither the perfect hcp crystal nor the
perfect bcc one show signals of supersolidity according to our PIMC results.

\section{Conclusions}

We have performed a microscopic study of the coexistence hcp-bcc line in
solid $^4$He using the path integral Monte Carlo method. This method allows
for a nearly ab-initio study where the only inputs are the mass and the
interatomic potentials. Our goal has been to get an accurate view of the
possible differences between both lattices along the measured coexistence
line. Apart from the intrinsic interest on the study of the bcc phase,
which has been scarcely studied in the past, we were stimulated by the
recent activity of Polturak and collaborators~\cite{eyal} who have shown intriguing
results using the TO technique.

Our results show a small influence of the lattice type when crossing the
coexistence line. The internal energies are similar and the differences
between both crystals are more attributable to the slight difference in
density than to the particular crystal symmetry. Also the radial
distribution functions show only very tiny differences. What is more
significant is the drop in the kinetic energy from the hcp phase to the bcc
one, that we have estimated to be 0.5 K, and that is in overall agreement
with experimental findings~\cite{blasdell}. The less compact structure of the bcc solid is
also observed in the results obtained for the density profiles and the
Lindemann ratios. Finally, we have calculated the one-body density matrix
in both sides of the coexistence line without observing any relevant
difference between them and, more importantly, without obtaining
off-diagonal long-range order which allowed us to conclude that a perfect
bcc crystal, at the temperatures where this phase is stable, is not a
supersolid. Work is in progress in our group to determine if the
introduction of point defects (vacancies) in the bcc crystal could induce
supersolidity at these apparently too high temperatures.

\section*{Acknowledgments}
The authors acknowledge partial financial support from the
DGI (Spain) Grant No.~FIS2008-04403 and Generalitat de Catalunya
Grant No.~2009SGR-1003.

\end{document}